\documentclass[]{aa}  
\usepackage{lineno} 
    \usepackage{graphicx}
    \usepackage{txfonts}
    \usepackage{booktabs}
    \usepackage[normalem]{ulem}
    \usepackage{multirow}
    \usepackage{floatrow}
    \usepackage[colorlinks = true, allcolors=blue]{hyperref}
    \usepackage{stfloats}
    \usepackage{booktabs}
    \usepackage{orcidlink}
    \usepackage{sidecap}
    \usepackage{pdflscape}
    \newcommand\farcsec{\mbox{$.\!\!^{\prime\prime}$}}%

    \usepackage[normalem]{ulem}
    \newcommand{\HII}{H{\sc ii~}}

\begin{document}

       \title{Star Formation Beyond the Optical Disk}
       \subtitle{The Low-Density Outskirts of NGC\,2090}
       \titlerunning{Star Formation in the extended disk of NGC\,2090}
       
       \author{Jyoti Yadav\orcidlink{0000-0002-5641-8102}\inst{1,2}
       \and
              Mousumi Das\orcidlink{0000-0001-8996-6474}\inst{3}
              \and
              S Amrutha\orcidlink{0009-0005-6072-9252}\inst{3}
              \and
              Dimitra Rigopoulou\orcidlink{0000-0001-6854-7545}\inst{4,5}
              }
    
       \institute{
            Instituto de Astrofísica de Canarias, Vía Láctea s/n, E-38205 La Laguna, Spain\\
            \email{yadavjyoti636@gmail.com}
            \and
            Departamento de Astrofísica, Universidad de La Laguna, E-38206 La Laguna, Spain\\
            \and
            Indian Institute of Astrophysics, Koramangala II Block, Bangalore 560034, India\\
            \and
            Oxford Astrophysics, Denys Wilkinson Building, University of Oxford, Keble Road, Oxford OX1 3RH, UK \\
            \and
            School of Sciences, European University Cyprus, Diogenes street, Engomi, 1516 Nicosia, Cyprus            
            }
    
    
     
      \abstract
       {We present a far-ultraviolet (FUV) analysis of the star-forming complexes (SFCs) in the nearby spiral galaxy NGC\,2090, based on observations from the Ultraviolet Imaging Telescope (UVIT), and compare it with emission from the optical and infrared bands. NGC\,2090 exhibits prominent star formation in its extended outer disk, with FUV emission traced out to $\sim$30 kpc, far beyond the truncation of the old stellar disk at $\sim$5 kpc. It is classified as an extended UV (XUV) disk galaxy. We identify and characterize the SFCs both within and beyond the optical radius (R$_{25}$), estimating their physical sizes and star formation rates (SFRs). The outer-disk SFCs are generally smaller in area and show a narrower distribution of SFR surface density ($\Sigma_{\mathrm{SFR}}$) compared to the inner-disk SFCs. We investigate the properties of the inner disk using mid-infrared data from the James Webb Space Telescope (JWST), and find that the polycyclic aromatic hydrocarbon (PAH) emission is strongly correlated with regions of active star formation. The specific SFR (sSFR) increases with radius, consistent with a scenario of inside-out disk growth. The observed number of SFCs and their H$\alpha$-to-FUV flux ratios in the outer disk of NGC\,2090 indicate ongoing massive star formation and are consistent with a top-heavy IMF, implying that the upper end of the IMF is not truncated in the low-density, metal-poor outskirts. These results suggest that XUV disks can host significant massive star formation despite their low stellar density and metallicity.
}
    
    
       \maketitle
    %

\section{Introduction}

Star formation in the outer disks of galaxies differs fundamentally from that in their inner regions, reflecting differences in stellar surface density and gas properties \citep{Bigiel2010AJ....140.1194B}. Inner disks typically have higher stellar and gas densities, higher metal content, and more dust, conditions that favor the formation of massive star-forming regions. In contrast, outer disks are more metal-poor, dust-poor, and have much lower stellar surface densities making them adverse environments for forming massive stars \citep{Bush2008ApJ...683L..13B, Bush2010ApJ...713..780B}. 

Despite the low stellar and gas densities observed in the outer disk, ongoing massive star formation has been observed beyond the optical radius (R$_{25}$) in several nearby galaxies through H$\alpha$ and UV emission \citep{Ferguson1998ApJ...506L..19F, Lelievre2000AJ....120.1306L, Cuillandre2001ApJ...554..190C, Blok2003MNRAS.341L..39D, Christlein2008ApJ...680.1053C, Herbert2010ApJ...715..902H, Goddard2010MNRAS.405.2791G, Werk2010AJ....139..279W, Yadav2021ApJ...914...54Y, Das2021JApA...42...85D}. A key breakthrough came with the discovery of extended ultraviolet (XUV) disks by the GALEX mission, which found that about 30\% of nearby galaxies (D < 40 Mpc) host significant star formation beyond R$_{25}$ \citep{Donas1981A&A....97L...7D, Paz2005ApJ...627L..29G, Paz2007ApJ...661..115G, Thilker2005ApJ...619L..79T, Thilker2007ApJS..173..538T}. These XUV disks are mainly seen in gas-rich, late-type spirals. There are mainly two types of XUV galaxies. Type-1 XUV galaxies show extended star formation following spiral structures into the outer disk, while Type-2 XUV galaxies show star formation in their low surface brightness (LSB) outer blue disks; some galaxies display both characteristics and are classified as mixed XUV disks. Other examples of star formation in diffuse environments are classical LSB galaxies, where molecular gas is rarely detected \citep{Das2006ApJ...651..853D, Das2010A&A...523A..63D}; however, far-ultraviolet (FUV) and H$\alpha$ observations reveal ongoing star formation activity \citep{Boissier2007ApJS..173..524B, Yadav2022A&A...657L..10Y}.

Cold gas, particularly molecular gas, is also critical for star formation. While outer disks of XUV galaxies show significant reservoirs of neutral atomic hydrogen (HI) \citep{Boissier2003MNRAS.346.1215B, Zaritsky2007AJ....134..135Z}, molecular gas traced by CO is usually scarce, and CO detections are rare \citep{Lundgren2004A&A...413..505L, Dessauges2014A&A...566A.147D, Bicalho2019A&A...623A..66B}.
Multicomponent stability analyses show that colder gas, characterized by a lower velocity dispersion, plays a dominant role in triggering disk instabilities and regulating the scales over which they develop \citep{Bertin1988A&A...195..105B, Romeo2013MNRAS.433.1389R, Romeo2015MNRAS.451.3107R, Romeo2016MNRAS.460.2360R, Romeo2017MNRAS.469..286R}. The relationship between HI gas and star formation has been extensively studied across a variety of galactic environments, including the outer disks of galaxies. While the general correlation remains similar to that observed in inner regions, it exhibits a noticeably steeper slope in the outskirts \citep{Bigiel2010AJ....140.1194B, Bigiel2010ApJ...720L..31B}. These regions are also characterized by significantly lower star formation efficiencies (SFE) \citep{Bigiel2010ApJ...720L..31B}, indicating that the physical conditions governing star formation in the outer disk may differ from those in the inner galaxy. 

Building on the role of gas in fueling star formation, galaxies are thought to continuously accrete material from beyond their virial radii over cosmic time, providing a sustained reservoir that supports ongoing star formation, particularly in their outer disks \citep{Rees1977MNRAS.179..541R, White1978MNRAS.183..341W, Katz1991ApJ...377..365K, Keres2005MNRAS.363....2K, Dekel2009Natur.457..451D}. Cosmological simulations predict that such cold gas can be transported inward at velocities of 20-60 km s$^{-1}$ from regions extending well beyond the stellar disk \citep{Dekel2006MNRAS.368....2D, Dekel2009Natur.457..451D}. Star formation in the extended disk can also be facilitated by the deposition of gas during recent or ongoing galactic interactions \citep{Thilker2007ApJS..173..538T, Yadav2023MNRAS.526..198Y}. In low-density environments, star formation is mainly attributed to external processes, including gas inflows induced by interactions or the gradual accretion of cold gas from the intergalactic medium or the cosmic web \citep{Lemonias2011ApJ...733...74L, Yadav2021A&A...651L...9Y, Yadav2024A&A...689A.346Y}.


Understanding the nature of star formation in galaxies requires knowledge of the stellar initial mass function (IMF), whose universality remains a central question across different galactic environments. Observational studies have long supported the idea of a universal IMF \citep{Kroupa2002Sci...295...82K, Bastian2010ARA&A..48..339B, Koda2012ApJ...749...20K}, yet growing evidence points to potential IMF variations driven by environmental factors such as gas density and metallicity. Systematic variations in the IMF are predicted by theoretical studies \citep{Adams1996ApJ...464..256A, Adams1996ApJ...468..586A, Larson1998MNRAS.301..569L, Dib2007MNRAS.381L..40D} and have been increasingly supported by observational evidence. A detailed review can be found in \citet{Kroupa2013pss5.book..115K, Hopkins2018PASA...35...39H}. 
High gas density, low-metallicity environments favours the formation of massive stars, resulting in a top-heavy IMF \citep{Matteucci1994A&A...288...57M, Dabringhausen2009MNRAS.394.1529D, Marks2012MNRAS.422.2246M, Zhang2018Natur.558..260Z, Schneider2018A&A...618A..73S, Kalari2018ApJ...857..132K, Brown2019ApJ...879...17B} and top light IMF is seen in low SFR systems \citep{Elmegreen2004MNRAS.354..367E, Meurer2009ApJ...695..765M, Lee2009ApJ...706..599L,Watts2018MNRAS.477.5554W}. Ultraviolet emission primarily traces massive O and B stars, while H$\alpha$ emission, originating from ionized gas, directly reflects the presence of the most massive O stars ($>$20 M$_\odot$) \citep{Kennicutt1998ARA&A..36..189K}. Therefore, the combined use of FUV and H$\alpha$ fluxes provides a powerful diagnostic to probe the upper end of the IMF, especially in extended, low-density outer disks of galaxies.

In this study, we investigate the star formation properties of the Type-2 XUV disk galaxy NGC\,2090, which exhibits prominent FUV emission well beyond its optical radius, indicative of ongoing star formation in its low-density outer regions. Our primary objective is to examine how star formation in the extended outer disks of Type-2 XUV disk galaxies compares with that in the bright inner disk. The extended disks have properties similar to those of  LSB galaxies. LSB galaxies generally exhibit low surface brightness, blue colors \citep{Blok1995MNRAS.274..235D}, substantial reservoirs of neutral gas \citep{Neil1998AJ....116.2776O}, and low metallicities \citep{McGaugh1994ApJ...426..135M}. Likewise, XUV regions are metal-poor \citep{Paz2007ApJ...661..115G} and are typically found in galaxies that are systematically more gas-rich than the general field population \citep{Thilker2007ApJS..173..538T}. These similarities suggest that star formation in XUV disks and LSB galaxies may share common properties. Thus, XUV galaxies provide us with a tool to understand star formation properties in low-density environments.


UV photons from massive young star-forming regions excite polycyclic aromatic hydrocarbons (PAHs), producing prominent emission features at 3.3, 6.2, 7.7, 8.6, and 11.3\,$\mu$m, thereby directly linking the emission to sites of active star formation \citep{Tielens2008ARA&A..46..289T, Li2020NatAs...4..339L}. 
In galaxies experiencing intense star formation, as much as 20\% of the total infrared luminosity is emitted in the PAH bands alone \citep{Smith2007ApJ...656..770S}. Observationally, they serve as effective tracers of both cold gas \citep{Rigopoulou1999AJ....118.2625R, Cortzen2019MNRAS.482.1618C, Gao2019ApJ...887..172G, Gao2022ApJ...940..133G, Leroy2023ApJ...944L...9L, Sandstrom2023ApJ...944L...8S, Whitcomb2023ApJ...948...88W, Chown2025ApJ...987...91C} and the heating processes associated with star formation activity \citep{Peeters2004ApJ...613..986P, Calzetti2007ApJ...666..870C, Calapa2014ApJ...784..130C, Cluver2017ApJ...850...68C, Pathak2024AJ....167...39P, Gregg2024ApJ...971..115G}. Studies have examined the effects of high-energy photons on dust in HII regions \citep{Riener2018A&A...612A..81R}, and recent James Webb Space Telescope (JWST) observations have begun to resolve the properties of PAHs in such regions with unprecedented detail \citep{Chastenet2023ApJ...944L..12C}. To investigate the characteristics of PAH molecules, it is essential to analyze high-resolution infrared observations. With the high sensitivity and spatial resolution in the mid-infrared, JWST now enables detailed studies of PAH emission in individual HII regions, supernova remnants, star clusters, and their immediate surroundings in nearby galaxies \citep{Dale2023ApJ...944L..23D, Egorov2023ApJ...944L..16E, Sutter2024ApJ...971..178S, Chastenet2023ApJ...944L..11C, Sandstrom2023ApJ...944L...7S, Pedrini2024ApJ...971...32P, Gregg2024ApJ...971..115G, Baron2024ApJ...968...24B, Baron2025ApJ...978..135B, Ujjwal2024A&A...684A..71U}. 
We therefore incorporate high-resolution mid-infrared imaging from the JWST, which provides a detailed view of the inner disk. This allows us to directly trace PAH emission surrounding young massive stars. By combining these JWST observations with other multi-wavelength diagnostics, we can robustly characterize star formation across diverse galactic environments and gain new insights into the star formation processes in NGC\,2090.

\section{NGC2090}
NGC\,2090 is an isolated spiral galaxy classified as SA:(rs)b. The UV emission is clearly detected in the extended outer disk, presenting as a smooth structural continuation of the main stellar disk rather than as discrete outer complexes or detached features. It is classified as a Type 2 XUV disk galaxy, as it has an extended outer LSB disk region that lies within the SF threshold, and is blue in color \citep{Thilker2007ApJS..173..538T}. The galaxy is at a redshift of 0.003075 \citep{Springob2005ApJS..160..149S} and has a log ${\rm M_{HI}}$ (M$_{\odot}$) $\sim$ 9.35 \citep{Thilker2007ApJS..173..538T}. NGC\,2090 hosts a significantly extended LSB zone, covering an area more than ten times larger than the effective radius (encompassing 80\% of the old stellar light) of the central stellar disk. Ultraviolet imaging from GALEX highlights a clear distinction between the inner and outer regions of the galaxy \citep{Thilker2007ApJS..173..538T}. The extended LSB zone shows numerous \HII regions, providing direct evidence of ongoing star formation in the extended outer disk \citep{Koopmann2006ApJS..162...97K}. The spiral arms in the outer disk appear to be flocculent in nature (fig.~\ref{fig:multiwavelength_images}). The details of NGC\,2090 are presented in Table~\ref{tab:details}.

\begin{table}
    \centering
    \begin{tabular}{cc}
    \toprule
    Source & NGC\,2090\\
    \hline
    R.A (J2000)&  05:47:01.8982\\
    Dec. (J2000)& -34:15:00.806 \\
    z  & 0.003075$\pm$2.00e-6  \\
    Distance  & $\sim$ 14.7 Mpc \\
    V$_{sys}$ & $\sim$ 922 kms$^{-1}$ \\
    D$_{25}$ & 4.3\arcmin \\
    HI mass  & 9.35$\times$10$^{10}$ M$_{\odot}$ \\
   log($\Sigma_{SFR}$(M$_\odot$ yr$^{-1}$ kpc$^{-2}$)) FUV &  -2.33 \\
    \toprule
    \end{tabular}
    \tablefoot{Details of NGC\,2090. Right ascension (R.A), declination (Dec.), redshift (z), distance and V$_{sys}$  are from NED; the optical diameter (D$_{25}$), the HI mass and log($\Sigma_{SFR}$(M$_\odot$ yr$^{-1}$ kpc$^{-2}$)) FUV are from \citet{Thilker2007ApJS..173..538T}.}
    \label{tab:details}
\end{table}

\section{Observations and analysis}\label{sec:observation}

We conducted deep FUV imaging of the galaxy NGC 2090 using the Ultraviolet Imaging Telescope (UVIT) onboard the AstroSat satellite \citep{Kumar2012}. UVIT consists of two co-aligned Ritchey-Chrétien telescopes: one dedicated to the FUV range (1300-1800 \AA) and the other designed for simultaneous imaging in the near-ultraviolet (NUV; 2000-3000 \AA) and visible bands. The instrument allows for simultaneous observations across all three channels, with the visible channel primarily utilized for drift correction. UVIT has a variety of photometric filters in both the FUV and NUV bands, features a field of view of approximately 28$\arcmin$, and achieves spatial resolutions of 1\farcsec4 in FUV and 1\farcsec2 in NUV. UVIT offers a spatial resolution that is approximately three times better than that of GALEX. 
We processed the Level 1 data using CCDLAB \citep{Joe2017}, a software with a graphical interface that applies corrections for field distortion, flat-field variations, and image drift to generate science-ready data. Astrometric calibration was carried out using Gaia data. 

We utilized archival imaging data from the JWST obtained with the Near Infrared Camera (NIRCam; \citealt{Rieke2005SPIE.5904....1R}) and the Mid-Infrared Instrument (MIRI; \citealt{Rieke2015PASP..127..584R}).
We have utilized archival JWST MIRI F770W and F2100W images and the NIRCAM filters F300M, F335M from onboard the JWST as the source of PAH emission (PI: Adam Leroy, Proposal ID: 3707). The F770W image has a resolution of 0.26\arcsec, and F2100W images have an angular resolution of 0.67\arcsec. NIRCAM filters F300M, F335M, and F360M have an approximate angular resolution of around 0.1\arcsec.

We utilized archival g, r, and z-band imaging data from the Dark Energy Camera Legacy Survey (DECaLS). DECaLS employs the Dark Energy Camera (DECam; \citealt{Flaugher2015AJ....150..150F}), which is mounted on the 4 meter Victor M. Blanco Telescope at the Cerro Tololo Inter-American Observatory. It covers a wide field of view of 2.2$^\circ$ in diameter and offers a pixel scale of 0.262 arcsec per pixel.

We use archival H$\alpha$ imaging data from \citet{Koopmann2006ApJS..162...97K}. The observations were conducted using the 0.9-meter telescope at Kitt Peak National Observatory, equipped with the Tek2K-1 CCD detector at the Cerro Tololo Inter-American Observatory. The images were processed by performing continuum subtraction, sky subtraction, and flux calibration. The resolution is $\sim$ 1$\arcsec$. A detailed description of the data acquisition and reduction procedures can be found in \citet{Koopmann2001ApJS..135..125K, Koopmann2006ApJS..162...97K}.

We additionally utilized archival data from the Spitzer Space Telescope, obtained with the Infrared Array Camera (IRAC). IRAC provides imaging in four mid-infrared bands. In this work, we use Channel 2 (4.5 µm) and Channel 4 (8.0 µm) observations.

\section{Analysis}

\subsection{Multiwavelength emission}
To investigate the star formation properties of NGC\,2090, we examine its emission across multiple wavelengths, as shown in Fig.~\ref{fig:multiwavelength_images}. NGC\,2090 is an actively star-forming galaxy with log SFR = -0.14 M$_\odot$ yr$^{-1}$ \citep{Thilker2007ApJS..173..538T}, and both the FUV and H$\alpha$ maps reveal significant ongoing massive star formation in its XUV disk. NGC\,2090 shows strong FUV and H$\alpha$ emission across both its inner and extended outer disk. However, the 2MASS K-band image shows dominant emission from the central region and no significant emission in the outer disk, consistent with an older, more evolved stellar population concentrated in the inner disk. The g-band image (central wavelength $\approx$ 473 nm) primarily traces the stellar continuum and shows the stellar disk (Fig.~\ref{fig:multiwavelength_images} top row 3rd column).

The JWST observations used in this study cover only the central regions of the NGC\,2090. The F335M filter captures both stellar continuum and PAH emission features. In particular, the prominent 3.3\,$\mu$m feature arises from the C--H stretching modes of small, predominantly neutral PAH molecules, which are commonly excited by ultraviolet photons in star-forming regions \citep{Ricca2012ApJ...754...75R, Draine2021ApJ...917....3D}. The F770W filter shows the emission from larger, ionized PAH grains, primarily through the C--C stretching and C--H in-plane bending modes \citep{Allamandola1989ApJS...71..733A, Galliano2008ApJ...679..310G, Maragkoudakis2018MNRAS.481.5370M, Maragkoudakis2022ApJ...931...38M, Draine2021ApJ...917....3D, Rigopoulou2021MNRAS.504.5287R}. The F2100W filter shows warm dust continuum emission. This emission originates from interstellar dust grains that absorb ultraviolet photons from young, massive stars and subsequently re-emit the energy in the mid-infrared. The F2100W image traces the locations of dust-embedded star-forming regions within the central regions of these galaxies. The JWST images, particularly in the F2100W band, display instrumental artifacts affecting certain regions of the data. These features likely arise from detector readout variations, residual background subtraction errors, or imperfect flat-fielding, and they may impact quantitative measurements.

\begin{figure*}
    \centering
    \includegraphics[width=0.43\linewidth]{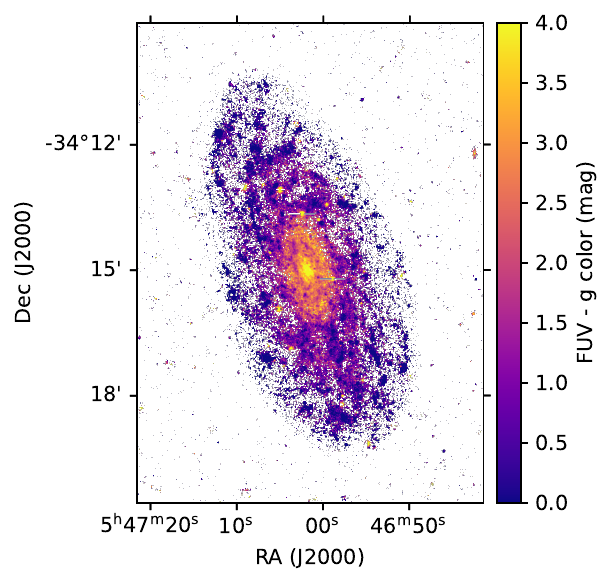}
    \includegraphics[width=0.45\linewidth]{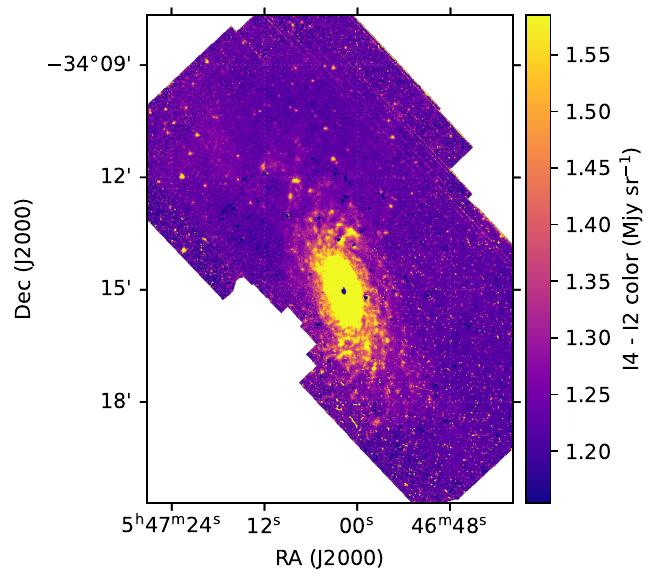}
    \caption{Left and right panels display the FUV-$g$ and Spitzer IRAC CH4-CH2 color maps of NGC,2090, respectively. The FUV–$g$ map highlights the bluer spiral arms, indicative of recent star formation, while the IRAC CH4-CH2 map traces regions of bright dust and PAH emission.}
    \label{fig:color_maps}
\end{figure*}
\subsubsection{Color maps}
Color maps of galaxies provide a powerful tool for tracing spatial variations in stellar populations, dust, and star formation history across the disk \citep{Mateos2007ApJ...658.1006M, Bakos2008ApJ...683L.103B}. In particular, FUV-g and IRAC CH4 (8.0\,$\mu$m) - CH2 (4.5\,$\mu$m) colors probe recent star formation and dust emission in galaxies, offering a complementary view of disk evolution \citep{Boissier2007ApJS..173..524B, Mateos2007ApJ...658.1006M, Paz2007ApJS..173..185G}. To explore the spatial distribution of stellar populations in NGC\,2090, we constructed two-dimensional color maps using our FUV, optical imaging, and Spitzer IRAC CH4 and CH2 data. We reprojected the g-band data to the pixel scale of the FUV imaging and corrected the Galactic extinction for each band. We convolved FUV imaging to g-band resolution and IRAC CH2 to CH4 resolution. Fig~\ref{fig:color_maps} shows the FUV-g (left) and IRAC CH4-CH2 (right). FUV-g map reveals that the spiral arms are significantly bluer than the inner disk, consistent with ongoing, recent star formation along the arms. However, the inner disk appears redder, which could be due to an older, more evolved population of cooler, low-mass stars concentrated in the central regions, or alternatively, stronger attenuation of FUV emission by dust, as the central regions typically have a higher dust content than the outer disk. 

The Spitzer IRAC CH4-CH2 color map is an effective tracer of PAH and dust emission. The CH4 band is dominated by strong PAH emission (7.7\,$\mu$m, 8.6\,$\mu$m) and hot dust, tracing star-forming regions, while CH2 traces mostly stellar continuum. The color map (Fig~\ref{fig:color_maps} right panel) is brighter along the spiral arms and within the central regions of the galaxy. These bright regions are indicative of localized dust lanes, where dust obscures the stellar light. 
Similarly, the dust content in the central regions can be associated with higher gas and dust densities. 

\subsubsection{SFCs detection and estimation of SFR}\label{sec:sfr_estimation}
\begin{figure*}
    \centering
    \includegraphics[width=0.95\textwidth]{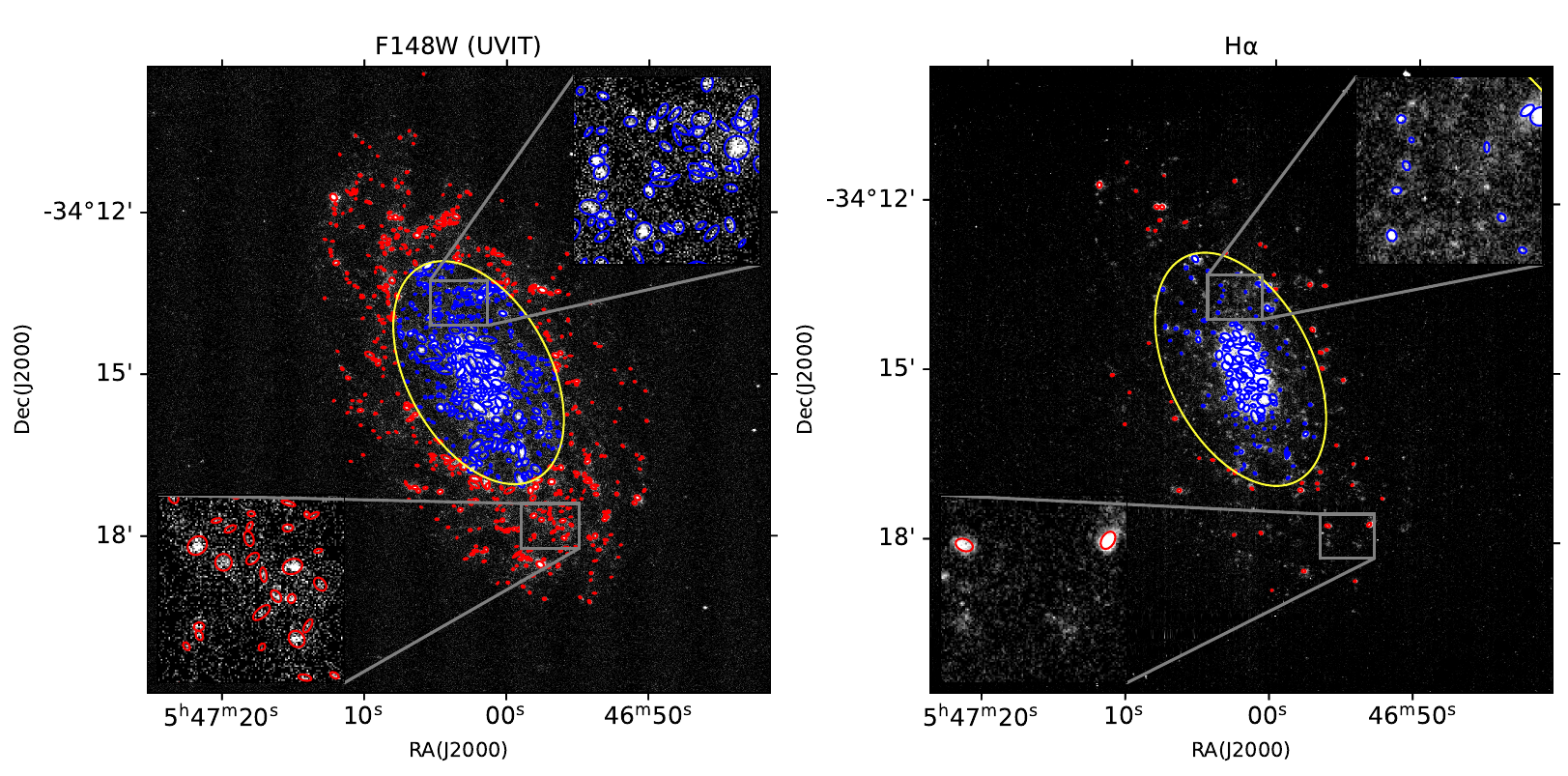}
    \caption{Identified SFCs in FUV(left) and H$\alpha$(right). Blue and red symbols denote SFCs located inside and outside the optical radius, respectively, while the yellow contour marks the optical radius. The grey boxes highlight the regions shown in the inset, providing a zoomed-in view of the SFCs within those areas.}
    \label{fig:identified_SFC}
\end{figure*}

NGC 2090 displays bright UV and H$\alpha$ emission extending throughout its outer disk, indicating the presence of star-forming complexes (SFCs). To identify and extract these SFCs, we employed the Source Extractor (SExtractor; \citealt{Bertin1996}), a Python-based tool widely used for source detection and photometry. SExtractor detects sources by applying a user-defined threshold, identifying connected pixels with flux values exceeding this limit. For our analysis, we adopted a detection threshold of 3$\sigma$, where $\sigma$ corresponds to the root mean square of global background noise.
Fig.~\ref{fig:identified_SFC} shows the identified SFCs in FUV(left) and H$\alpha$(right). We divided the SFCs as inner and outer based on the optical radius (R$_{25}$).

We performed elliptical aperture photometry on the detected sources using the Photutils package. Foreground Galactic extinction was corrected using the Fitzpatrick extinction law \citep{fitzpatrick1999}, assuming a total-to-selective extinction ratio of R(V) = 3.1.

\begin{equation}
A_{\lambda}  =R_{\lambda}\times E(B-V)
\end{equation}
where $A_{\lambda}$ is the extinction at wavelength $\lambda$, and E(B-V) is the reddening.

The NUV data from UVIT are not available; thus, we used GALEX FUV and NUV data for internal extinction correction. We estimated the UV $\beta$ slope using GALEX FUV and NUV data using the relation provided by \citet{Onodera2016ApJ...822...42O}.
\begin{equation}
    \beta_{UV} = -0.4 \frac{(m_{FUV} - m_{NUV}))}{log(\lambda_{FUV}/\lambda_{NUV})} - 2
\end{equation}

We used the $\beta_{UV}$ to estimate the FUV extinction using the relation provided by \citet{Figueira2022A&A...667A..29F}.
\begin{equation}
A_{FUV} = 1.78 \beta_{FUV} + 3.71 
\end{equation}
We created an FUV extinction map from GALEX and reprojected it to UVIT pixel scale, assuming that the extinction does not vary much within GALEX resolution.


To correct for attenuation by dust within the galaxy in H$\alpha$ we used the correlation A$_{H\alpha}$ = A$_{FUV}$/ 3.6 \citep{Leroy2008AJ....136.2782L} and reprojected the extinction map to H$\alpha$ image.

We calculated the SFR in FUV for each of the SFCs using the following formula \citep{Leroy2012AJ....144....3L, Salim2007}:
\begin{equation}
\label{eqn:sfruv}
SFR_{(FUV)}  = 0.68\times10^{-28} L_{FUV}
\end{equation} 
Where SFR$_{(FUV)}$ is the SFR in FUV [M$_\odot$ yr$^{-1}$] and $L_{FUV}$ is the FUV Luminosity [erg s$^{-1}$ Hz$^{-1}$].

We used \citet{Calzetti2007ApJ...666..870C} law to calculate the SFR for each SFC in H$\alpha$.
\begin{equation}
\label{eqn:sfrha}
SFR_{(H\alpha)}  = 5.3\times10^{-42} L_{H\alpha}
\end{equation} 
Where SFR$_{(H\alpha)}$ is the SFR in H$\alpha$ [M$_\odot$ yr$^{-1}$] and L$_{H\alpha}$ is the H$\alpha$ Luminosity [erg s$^{-1}$]. We define $\Sigma_{SFR}$ as SFR/Area of the SFCs.


\begin{figure*}
    \centering
    \includegraphics[width=0.85\textwidth]{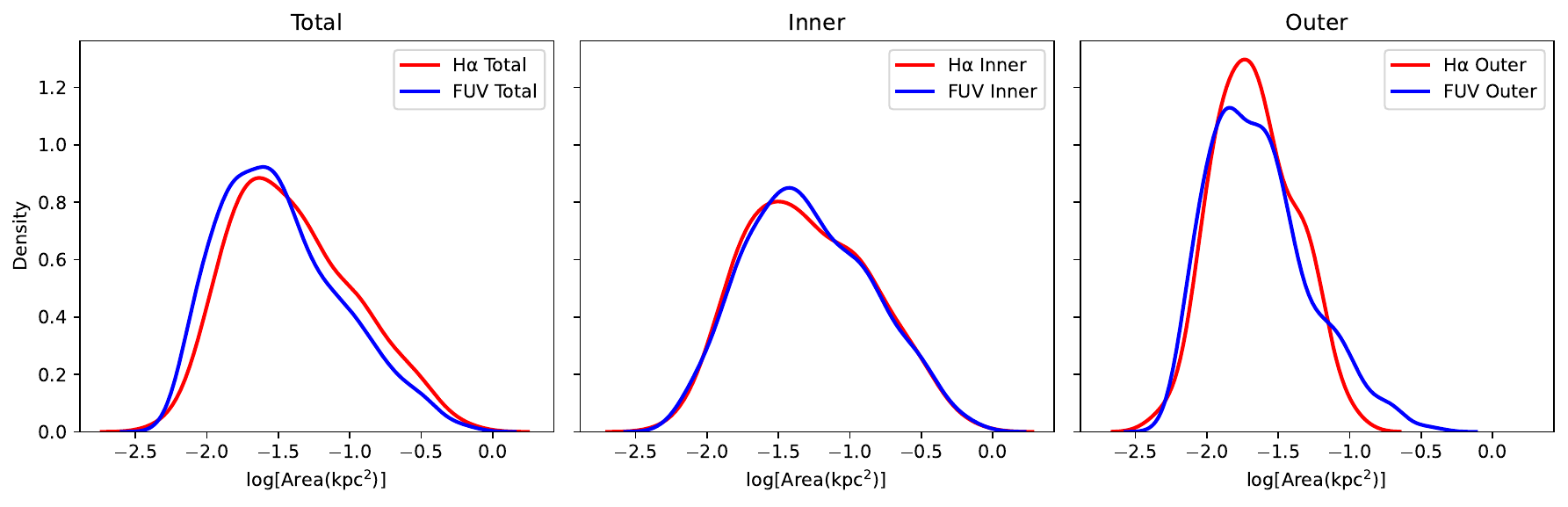}
    \includegraphics[width=0.85\textwidth]{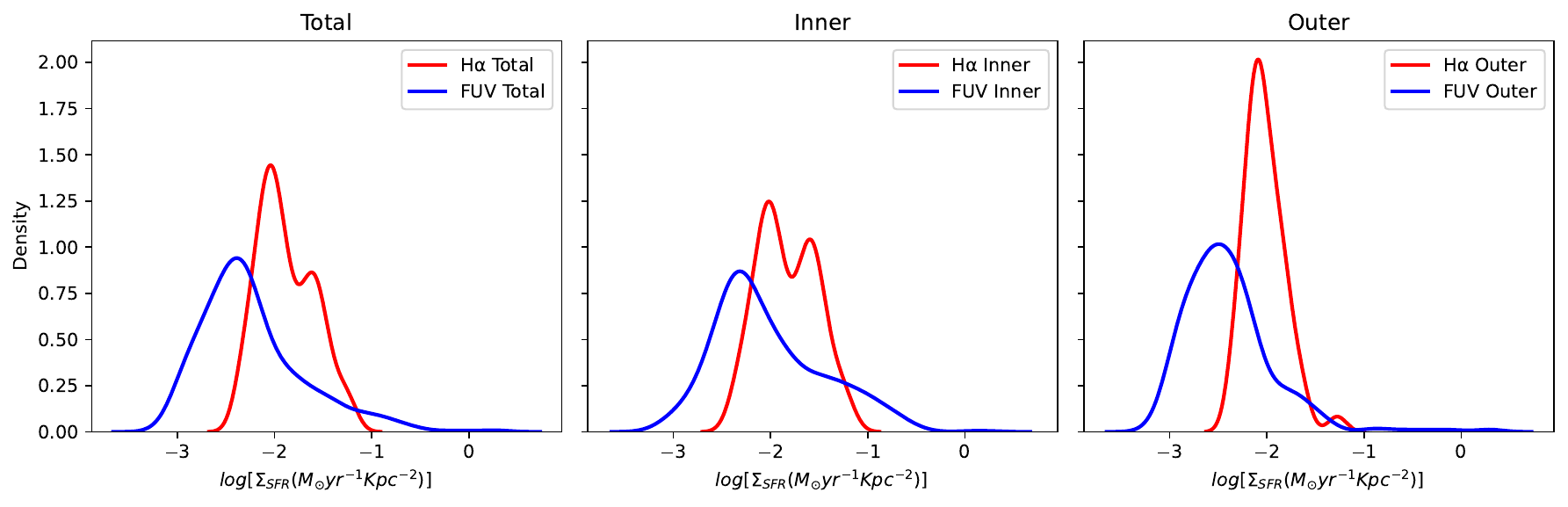}
    \caption{The top and bottom panels show the area and $\Sigma_{SFR}$ of SFC's in FUV (blue) and H$\alpha$ (red), respectively.}
    \label{fig:all_sfcs}
\end{figure*}

The number of identified SFCs in FUV and H$\alpha$ is given in Table \ref{tab:sfcs}. The FUV has a higher number of SFCs in the outer disk compared to the inner disk, while H$\alpha$ emission predominantly traces SFCs in the inner disk. The number of SFCs detected in FUV is larger than that identified in H$\alpha$ in both regions, indicating more spatially extended and longer-timescale star formation that is preferentially traced by FUV emission.

The detection of H$\alpha$ emission in the outer disk traces ionized gas powered by massive O-type stars and therefore implies star formation on timescales of only a few Myr \citep{Kennicutt1998ARA&A..36..189K}. This supports the existence of recent, high-mass star formation in the low-density outer regions. The presence of such stars argues against a truncated IMF in the outer disk. A truncated IMF would fail to produce the most massive stars altogether, while a stochastic IMF could still form them occasionally, though with low probability in low-mass clusters. Thus, detecting even a small number of outer disk H$\alpha$ sources implies that the upper IMF is not truncated.

\begin{table}[]
    \centering
    \begin{tabular}{c|c|c}
    Region & FUV & H$\alpha$ \\ 
     \hline
     Inner & 360 & 144 \\
     Outer & 453 & 52  
    \end{tabular}
    \caption{Number of detected SFCs in in FUV and H$\alpha$ in the inner and outer disks.}
    \label{tab:sfcs}
\end{table}

\citet{Koda2012ApJ...749...20K} demonstrated that, under the assumptions of instantaneous cluster formation and a constant cluster formation rate, the ratio of H$\alpha$ emitting to FUV emitting clusters reflects the ratio of their characteristic lifetimes, such that $N_{\mathrm{H}\alpha}/N_{\mathrm{FUV}} = t_{\mathrm{H}\alpha}/t_{\mathrm{FUV}}$. FUV emission traces star formation over timescales of approximately 100 Myr \citep{Donas1984A&A...140..325D, Schmitt2006ApJ...643..173S, Thilker2007ApJS..173..538T}, while H$\alpha$ emission is sensitive to more massive and younger stellar populations, up to around 10 Myr \citep{Weisz2012ApJ...744...44W, Caplar2019MNRAS.487.3845C, Haydon2020MNRAS.498..235H}. We measure this number ratio in the outer disk of NGC\,2090 to be $N_{\mathrm{H}\alpha}/N_{\mathrm{FUV}} \approx 0.11$, similar to the value reported by \citet{Koda2012ApJ...749...20K} for M83. Their models using \textsc{Starburst99}, a Salpeter IMF with $(M_l, M_u) = (0.1, 100)$\,M$_\odot$ and subsolar metallicity (0.2\,Z$_\odot$) predict $t_{\mathrm{H}\alpha}/t_{\mathrm{FUV}} \sim 0.08$, and they measured an observational ratio of $N_{\mathrm{H}\alpha}/N_{\mathrm{FUV}} = 0.10 \pm 0.03$. Our results are thus consistent with a standard Salpeter IMF in the outer disk.

Figure~\ref{fig:all_sfcs} shows the distribution of area and  $\log(\Sigma_{\mathrm{SFR}}$) , for the identified SFCs in both FUV and H$\alpha$. The physical sizes of SFCs are comparable between the two tracers across the inner and outer disks. However, the inner disk SFCs span a wider range in area compared to those in the outer disk. The inner disk SFCs have log(area(kpc$^{2}$)) values ranging from -2 to 0, whereas the outer-disk SFCs range from -2 to -1. The $\log(\Sigma_{\mathrm{SFR}}$(M$_{\odot}$ yr$^{-1}$ kpc$^{-2}$)) ranges from approximately $-3$ to $-1$ in FUV and from $-2.5$ to $0$ in H$\alpha$. 
The H$\alpha$ data from \citet{Koopmann2006ApJS..162...97K} have not been corrected for [N\,\textsc{ii}] contamination ($\lambda\lambda\,6548.1,\,6583.8$\,\AA). \citet{Kennicutt1992ApJ...388..310K} estimated a median [N\,\textsc{ii}]/H$\alpha$ ratio of 0.53 in H\,\textsc{ii} regions, which decreases in fainter galaxies \citep{Jansen2000ApJS..126..331J}.  \citet{James2005A&A...429..851J} further demonstrate that the [N\,\textsc{ii}]/H$\alpha$ ratio is lower than previously estimated and shows a strong radial dependence, indicating that the application of a single correction factor to all SFCs is not appropriate. Consequently, the H$\alpha$ images contain contributions from [N\,\textsc{ii}] emission, which likely leads to systematically higher $\Sigma_{\mathrm{SFR}}$ estimates compared to those derived from FUV.

\subsection{PAH Emission in the JWST F335M Band}
In most AKARI/IRC studies, the strength of the 3.3\,$\mu$m PAH feature is estimated by fitting a linear continuum and integrating the emission above it \citep{Imanishi2008PASJ...60S.489I, Imanishi2010ApJ...721.1233I, Ichikawa2014ApJ...794..139I, Inami2018A&A...617A.130I, Lai2020ApJ...905...55L}. We also use the similar method using SPITZER IRAC1. The SPITZER IRAC1 band broadly overlaps with the JWST F335M filter, both covering emission near 3.6\,$\mu$m. To estimate the continuum shape in this regime, we first reprojected and convolved the JWST F335M and F330M images onto the IRAC1 pixel grid and resolution, respectively. We then subtracted the F335M image from the IRAC1 map, producing a residual image we refer to as F$\arcmin$360, which approximates the continuum near 3.6\,$\mu$m. 

 Using the data from F300M and F$\arcmin$360, we derived the continuum slope using the following relation:

\begin{equation}
    m\arcmin = \frac{F\arcmin360-F300M}{\lambda_{eff F360}\arcmin-\lambda_{eff F300}}
\end{equation}

where F$\arcmin$360 and F300M represents the data and $\lambda_{effF360M}$ and $\lambda_{effF300M}$ are the effective wavelengths of the respective filters.
We reprojected the slope map, denoted as m$\arcmin$, onto the F335M pixel grid and refer to the resulting reprojected slope as m. We  used this slope to estimate the continuum contribution at the location of the F335M band, using the following equation:

\begin{equation}
    \mathrm{F335M\ Cont} = m \left[\lambda_{\mathrm{eff F335M}} - \lambda_{\mathrm{eff F300}}\right] + \mathrm{F300M}
\end{equation}

\begin{figure*}
    \centering
    \includegraphics[width=0.9\textwidth]{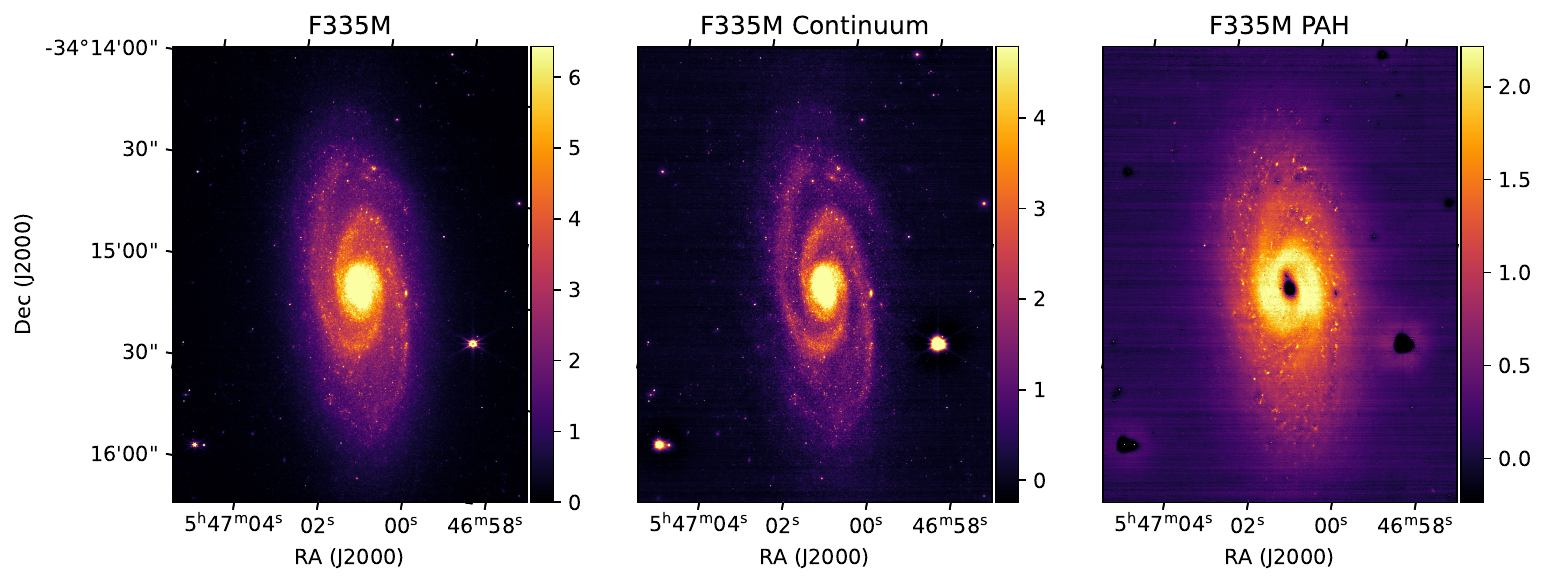}
    \caption{The left, middle, and right panels show the observed F335M image, the underlying stellar continuum in F335M, and the isolated PAH emission component at 3.3\,$\mu$m, respectively.}
    \label{fig:330PAH}
\end{figure*}

We subtracted the continuum F335M$_{Cont}$ from F335M filter data to get the F335M$_{PAH}$ emission. Figure~\ref{fig:330PAH} shows the emission maps in the F335M filter: the observed F335M image (left panel), the underlying stellar continuum component F335M$_{Cont}$ (middle panel), and the isolated 3.3\,$\mu$m PAH emission component F335M$_{PAH}$(right panel). The x-y axes of the maps are not aligned with the RA-Dec coordinate system, as the coordinate transformation alters both flux and pixel values. To preserve the integrity of the measurements, we retain the original orientation. The apparent central deficit in PAH emission is most likely not physical but instead arises from uncertainties in the continuum subtraction process. We therefore interpret the central PAH cavity with caution and do not consider it robust evidence for a true depletion of PAH emission in the galaxy center. The F335M$_{PAH}$ map reveals a clumpy morphology with compact regions distributed along the spiral arms of NGC\,2090. These clumps spatially coincide with regions of FUV emission, which trace young, massive SFCs. The correlation between the FUV bright SFCs and the enhanced 3.3\,$\mu$m PAH emission strongly suggests that the ultraviolet photons emitted by these young stars excite the small PAH molecules responsible for the 3.3\,$\mu$m feature. 

\begin{figure*}
    \centering
    \includegraphics[width=0.75\linewidth]{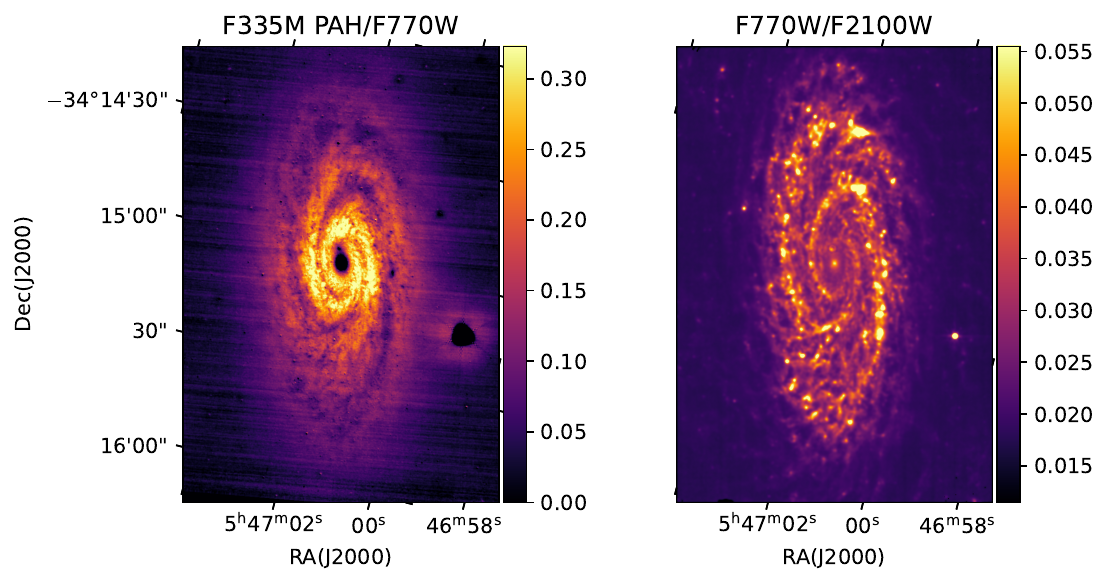}
    \caption{The left, and right panels show the maps of the F335M$_{PAH}$/F770W and F770W/F2100W band ratios, respectively.
    }
    \label{fig:PAH_ratio}
\end{figure*}

\subsection{Band Ratios}
Mid-infrared flux ratios using JWST bands provide powerful diagnostics of dust properties and PAH  emission in star-forming regions. Infrared emission features at 3.3, 6.2, 7.7, 8.6, and 11.3\,$\mu$m are commonly attributed to infrared fluorescence from PAH molecules primarily excited by FUV photons, with additional contributions from lower-energy NUV and optical radiation. As a result, these features broadly trace the recent star formation \citep{Rigopoulou1999AJ....118.2625R, Peeters2004ApJ...613..986P}. Ratios such as F335/F770 and F770/F2100 provide a means to compare the relative contributions of PAH emission and hot dust continuum, and to examine variations associated with star-forming regions \citep{Smith2007ApJ...656..770S, Yamada2013PASJ...65..103Y, Rigopoulou2024MNRAS.532.1598R, Gregg2024ApJ...971..115G}
These diagnostics are crucial for understanding the interplay between star formation, dust heating, and ISM conditions in galaxies.

To investigate the spatial variations of PAH emission and dust continuum in NGC\,2090, we constructed band ratio maps using F335M/F770W and F770W/F2100W. We reprojected the F335M image to align with the pixel grid of the F770W image, while the F770W and F2100W images from MIRI share the same native pixel scale. We convolved F335PAH to F770W resolution to get F335PAH/F770W. We convolved F770W to F2100W resolution to get the F770W/F2100W band ratio.
The resulting ratio maps display substantial spatial variations. 
Although the JWST artifacts are present in the maps, they do not significantly affect the derived ratio maps. The spatial variations in the emission ratios remain clearly discernible across the galaxy. The 7.7\,$\mu$m emission is consistently higher than the 3.3\,$\mu$m emission throughout the inner disk. However, the F335M/F770W ratio shows localized enhancements along the spiral arms, indicating a relative increase in the 3.3\,$\mu$m emission with respect to the 7.7\,$\mu$m feature in these regions, despite the latter remaining dominant. This suggests that the spiral arms host physical conditions favorable for the excitation of the 3.3\,$\mu$m feature, which is typically associated with small, neutral PAHs excited by UV photons from young stars. The elevated F335M/F770W ratios may therefore reflect a higher relative contribution from small PAHs in star-forming complexes. Incomplete subtraction of the stellar and hot dust continuum can artificially reduce the PAH signal in the nucleus, producing an apparent central depletion in the F335M/F770W ratio that is likely not real. This suggests that the inner PAH distribution should be interpreted with caution and may not reflect a true physical deficit. In contrast, the overall high 7.7\,$\mu$m emission in the inner disk is consistent with an increased fraction of ionized PAHs, as commonly observed near active star formation sites or active galactic nuclei. \citep{Egorov2023ApJ...944L..16E, Rigopoulou2024MNRAS.532.1598R, Garcia2024A&A...681L...7G, Zhang2025ApJS..280...65Z}. Although theoretical studies suggest that small PAHs are more easily destroyed in harsh environments with hot gas and strong radiation fields \citep{Micelotta2010A&A...510A..36M}, it is also possible that the harder radiation near SFCs primarily heats the PAHs, shifting their emission to shorter wavelengths and enhancing the 3.3\,$\mu$m feature without a significant change in the small-PAH fraction \citep{Draine2021ApJ...917....3D}.

The F770W/F2100W ratio shows the PAH strength relative to warm dust continuum. The F770W/F2100W ratio maps show relatively higher values in regions corresponding to SFCs, indicating that PAH emission is relatively stronger compared to warm dust emission. This trend is consistent with findings by \cite{Dale2025AJ....169..133D}, who reported a positive correlation between the age of compact stellar clusters and the degree of PAH ionization. 

\begin{figure*}
    \centering
    \includegraphics[width=0.35\linewidth]{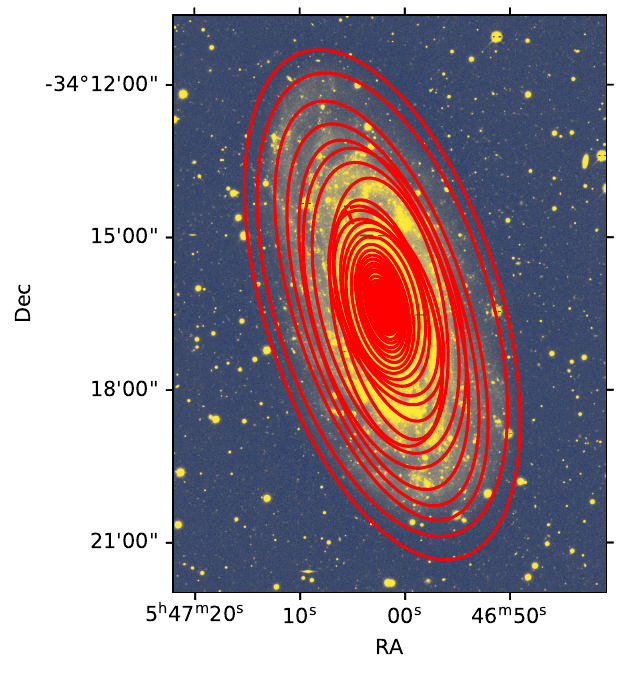}
    \includegraphics[width=0.40\linewidth]{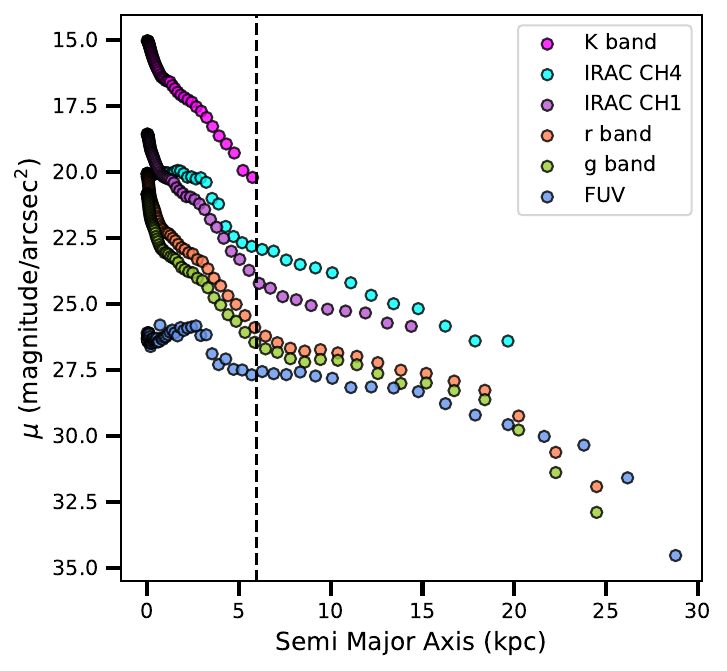}
    \caption{The left panel shows the elliptical isophotal fits on the g-band image, right panel shows the multiwavelength radial surface brightness profiles of NGC\,2090. The dashed vertical line marks the radial extent of the galaxy as traced by the K-band emission.
    }
    \label{fig:radial_profile}
\end{figure*}

\begin{figure*}
    \centering
    \includegraphics[width=0.8\linewidth]{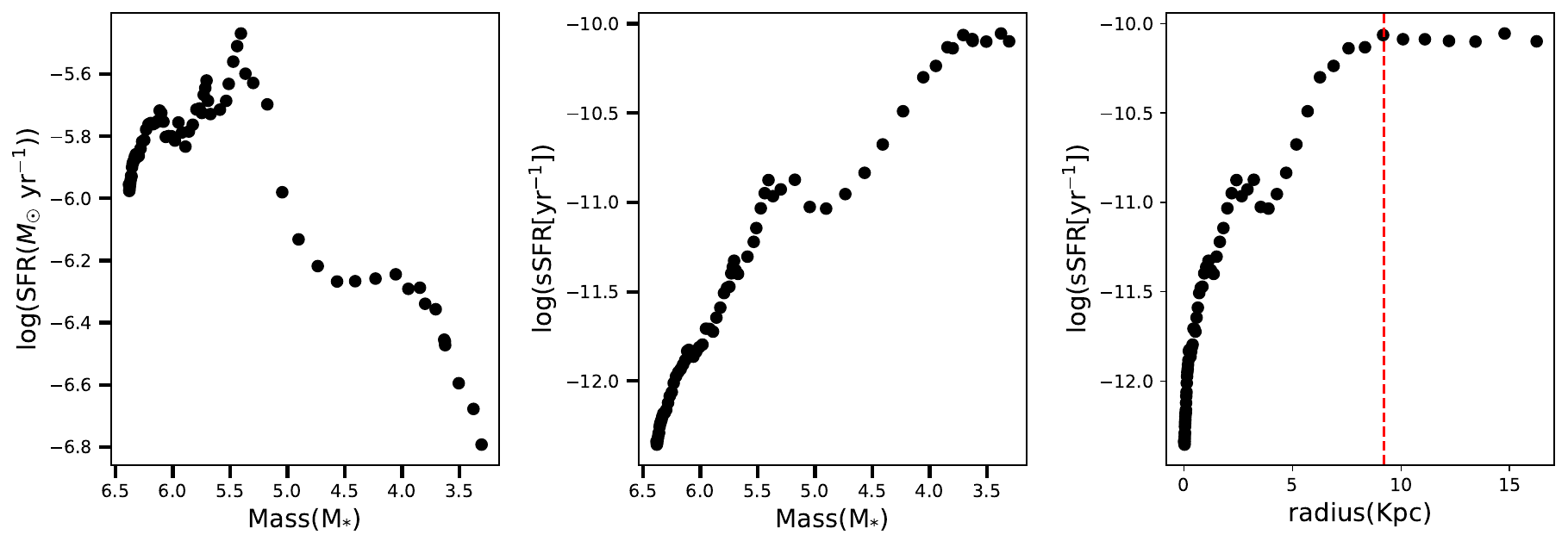}
    \caption{The left panel shows the star formation rate as a function of M$_{*}$. The middle panel shows the sSFR versus M$_{*}$. The right panel shows the radial profile of sSFR. The red dashed line shows the R$_{25}$. The x-axis in the left and middle panels corresponds to increasing radius.}
    \label{fig:SFR_profile}
\end{figure*}

\subsection{Radial Profiles}
\subsubsection{Multiwavelength Analysis}
We used the Photutils package to perform elliptical isophote fitting and derive the radial surface brightness profiles of the galaxy. The isophotal fitting was carried out using an iterative approach based on the algorithm introduced by \citet{Jedrzejewski1987MNRAS.226..747J}, which allows for accurate modeling of the light distribution. This method fits a series of elliptical isophotes to the galaxy image, adjusting parameters such as the semi-major axis length, ellipticity, and position angle at each step to best match the observed flux distribution. The resulting radial profile provides insight into the structural properties of the galaxy, including variations in surface brightness, geometry, and potential breaks or transitions in disk morphology. This approach is particularly effective in tracing the radial variation in surface brightness, enabling identification of transitions from high-intensity inner regions to the low surface brightness outskirts. It provides a robust framework for analyzing structural changes and detecting extended star-forming extended disks. Such detailed multiwavelength radial profile analysis is essential for understanding structural changes in the disk and identifying regions of extended star formation activity.

Figure~\ref{fig:radial_profile} presents the radial surface brightness profiles of NGC\,2090 across multiple wavelengths. The inner disk shows higher brightness in the longer-wavelength bands, such as the K-band, consistent with a centrally concentrated, older stellar population. The K band profile exhibits a truncation around 5kpc, highlighting the limited radial extent of the evolved stellar component. In contrast, the FUV profile extends significantly farther, reaching out to ~30kpc, whereas the intermediate-wavelength bands (e.g., optical g-band) typically truncate around 25kpc. Thus, the FUV emission is the most extended and indicates ongoing or recent star formation in the outer disk. It supports the inside-out formation scenario of the the stellar disk in which galaxies build up their stellar disks by forming new stars preferentially at larger radii over time.

Another interesting observation from Figure~\ref{fig:radial_profile} is that the Spitzer band 4 continuum emission extends much further out compared to the K band profile which represents old stars. It also extends beyond the stellar profile. Band 4 emission arises from dust and is possibly associated with recent star formation. So its larger radial extent also reinforces the idea that the XUV disk represents the inside-out growth of the stellar disk, which is possibly driven by gas accretion in the outer disk.

\subsubsection{SFR and sSFR Profile}
A key signature of inside-out disk growth is the radial variation of SFR and sSFR across galactic disks. In extended XUV disk galaxies, SFR and sSFR profiles often reveal enhanced star formation activity in the outer regions, beyond the traditional optical radius \citep{Thilker2007ApJS..173..538T, Mateos2007ApJ...658.1006M}. To examine how star formation varies from the inner to the outer regions of NGC\,2090, 
we used the radial profile information obtained from elliptical isophote fitting. We derive the radial profiles of the FUV and IRAC Channel 1 surface brightness using elliptical isophote fitting, which are subsequently converted into SFR and stellar mass (M$_{*}$), respectively. The SFRs were derived from the FUV emission profiles, while the stellar masses were estimated from the IRAC CH1 (3.6\,$\mu$m) band fitting using the mass--luminosity relation $\Upsilon = 0.35,\mathrm{M_{\odot},L_{\odot}^{-1}}$ from \citet{Schombert2022AJ....163..154S}. The variation of SFR and sSFR as a function of M$_{*}$, is shown in Fig.~\ref{fig:SFR_profile}. The SFR profile decreases with stellar mass and exhibits distinct peaks at specific radii, coinciding with the locations of the spiral arms. These peaks indicate enhanced star formation activity, likely driven by the elevated gas densities and favourable conditions along the spiral structure. Both the average SFR and stellar mass show a steady decline across the outer disk.

The sSFR profile as a function of stellar mass reveals that sSFR increases toward lower average stellar mass regions in the outer disk. The sSFR in the outer disk is systematically higher compared to the inner regions, shown in the right panel of Fig.~\ref{fig:SFR_profile}. This suggests that, although the outer disk hosts less average total stellar mass, it is currently experiencing relatively more active star formation per unit mass. This is also evident from the sSFR-radius plot. The sSFR increases from inner to outer disk and saturates to a constant value in the XUV disk at around $\sim$ 8 kpc.  The higher sSFR in the outer disk is consistent with the inside-out formation scenario, in which galaxies build up their stellar disks over time by forming new stars preferentially at larger radii.
The growth of local star-forming galaxies has been observed on spatially resolved scales, showing that galaxies generally grow inside-out \citep{Munoz2007ApJ...658.1006M, Perez2013ApJ...764L...1P, Gonzalez2015A&A...581A.103G, Pezzulli2015MNRAS.451.2324P, Garcia2017A&A...608A..27G, Frankel2019ApJ...884...99F, Amrutha2024MNRAS.530.2199A}. This result highlights the growth of the outer disk of NGC\,2090 driven by recent star formation.

\begin{figure*}
    \centering
    \includegraphics[width=0.45\linewidth]{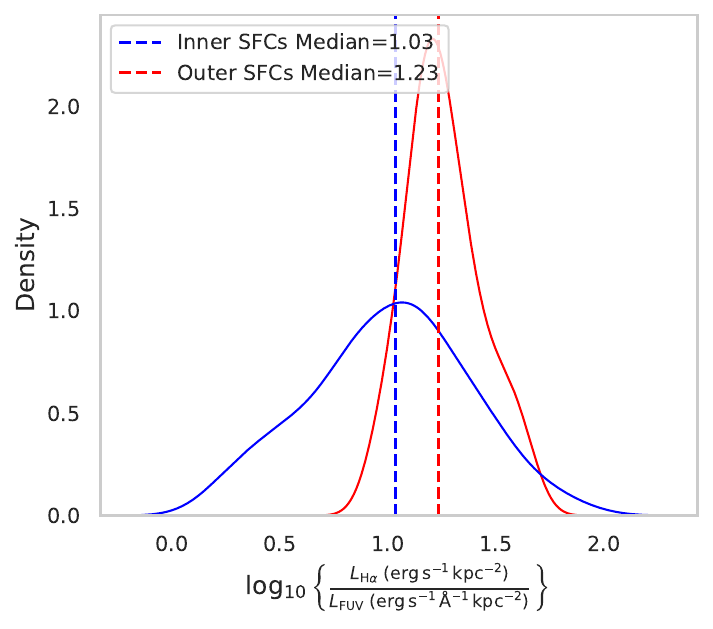}
    \includegraphics[width=0.4\linewidth]{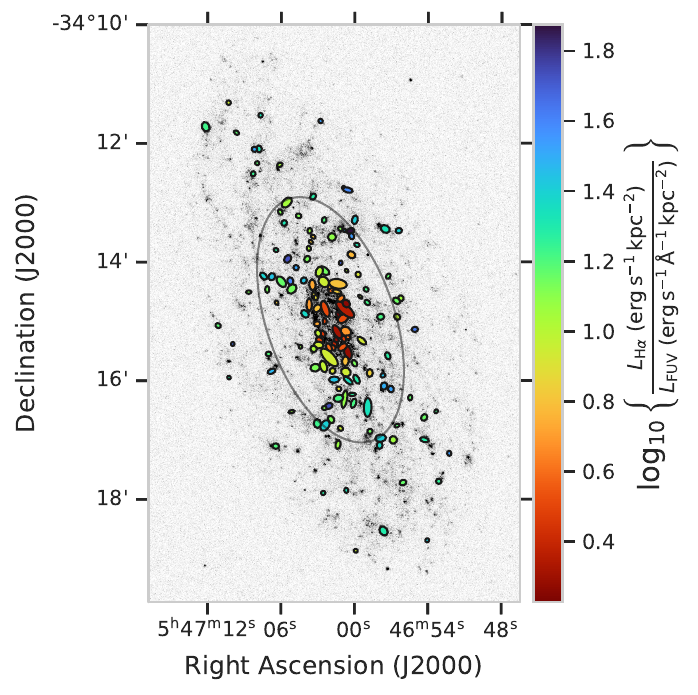}
    \caption{The left panel presents the histogram of the H$\alpha$-to-FUV flux ratio ({$F_{\rm H\alpha}/f_{\rm \lambda,FUV}$}) for SFCs in the inner and outer disk. The right panel shows the spatial variation of this flux ratio over the FUV map, where the cross-matched SFCs are plotted with symbol sizes proportional to their areas and colors representing the corresponding flux ratios.}
    \label{fig:flux_ratio}
\end{figure*}

\begin{figure*}
    \centering
    \includegraphics[width=0.42\linewidth]{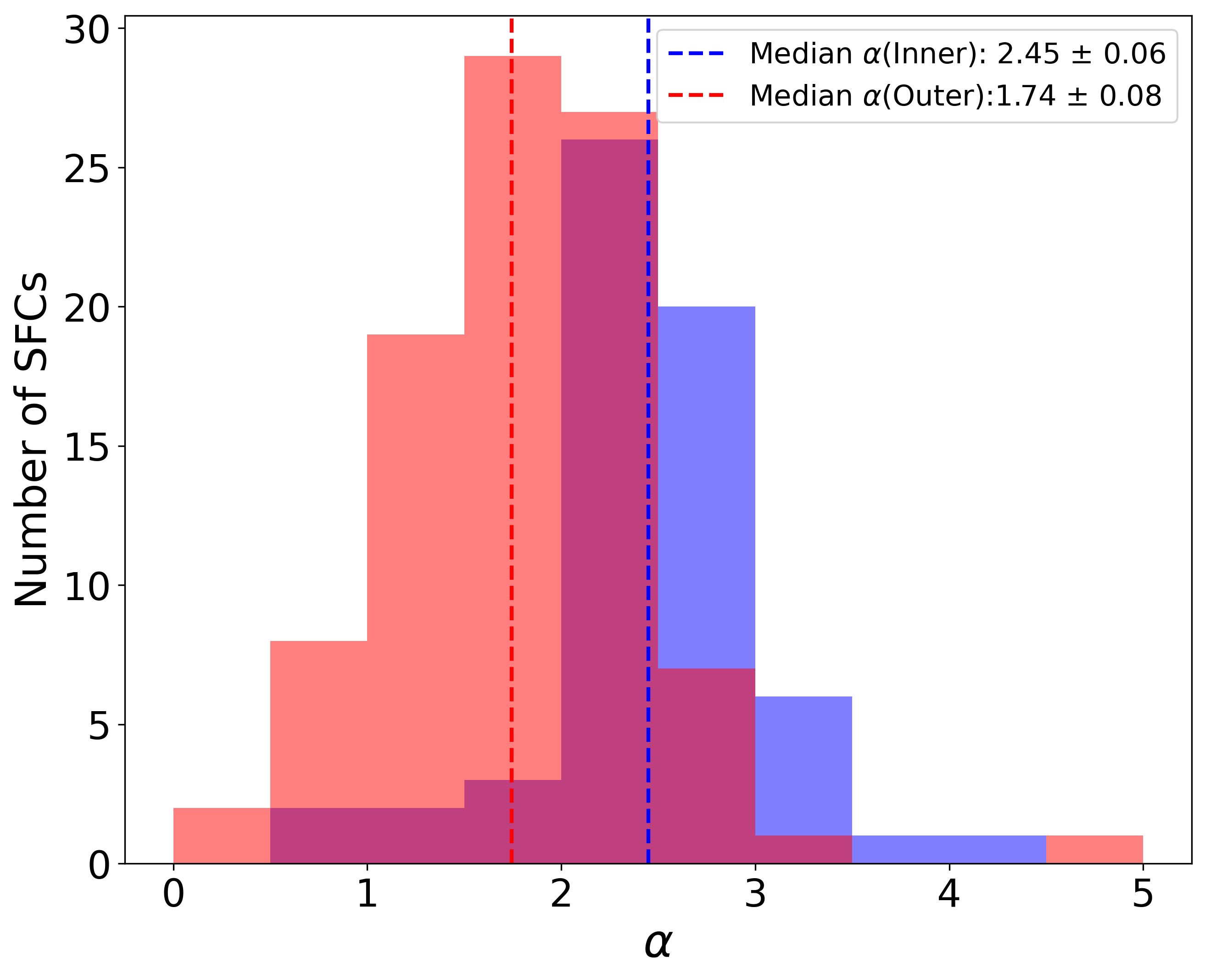}
    \includegraphics[width=0.45\linewidth]{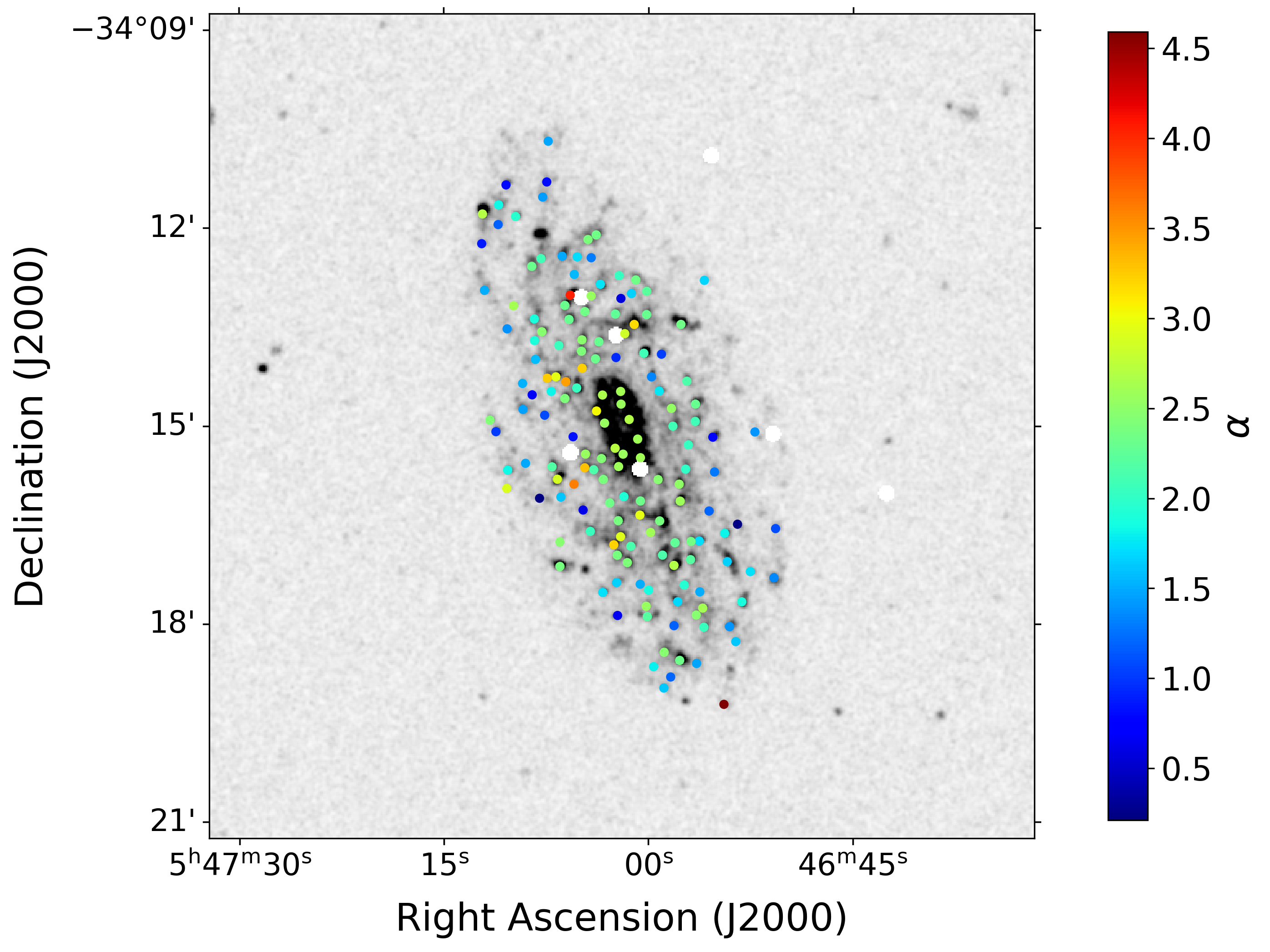}
    \caption{Distribution of IMF index $\alpha$ of SFCs over the disk of NGC 2090. The left panel presents the histogram of the $\alpha$ for SFCs in the inner and outer disk. The right panel shows the spatial variation of $\alpha$ over the FUV map.}
    \label{fig:IMF}
\end{figure*}

\section{IMF Variation}
The H$\alpha$-to-FUV flux ratio ($F_{\mathrm{H}\alpha}/f_{\lambda,\mathrm{FUV}}$) is used as a diagnostic for probing the upper end of the stellar IMF and recent star formation activity in galaxies \citep{Rautio2024A&A...681A..76R}. In particular, when applied to XUV disks, this ratio provides insight into whether the IMF or the mode of star formation changes in the low-density, outer disk environments. 

In our analysis of NGC\,2090, common SFCs in the H$\alpha$ and FUV images were identified by cross-matching their central positions within a 2\arcsec\ radius. This radius was selected based on the $\sim$1\arcsec\ spatial resolution of both datasets and the extended sizes of the SFCs. A smaller matching radius may miss true counterparts due to positional offsets within individual complexes, whereas a larger radius would increase the likelihood of chance matches. For each matched SFC, we calculated the H$\alpha$-to-FUV flux ratio in both the inner and outer disks. Figure~\ref{fig:flux_ratio} presents the distribution of $\log(F_{\mathrm{H}\alpha}/f_{\lambda,\mathrm{FUV}})$ for the inner and outer regions.

We find that the outer disk exhibits a median flux ratio of $\log(F_{\mathrm{H}\alpha}/f_{\lambda,\mathrm{FUV}})$ $\approx$ 1.23, while the inner disk shows a value of $\approx$ 1.03, albeit with a larger scatter ((Figure~\ref{fig:flux_ratio} left panel)). The ratio increases from inner to outer disk (Figure~\ref{fig:flux_ratio} right panel). \citet{Rautio2024A&A...681A..76R} employed \textsc{Starburst99} simulations to model this ratio for a metallicity of 0.2\,Z$_\odot$ and an upper stellar mass limit of $M_{\mathrm{u}} = 100$\,M$_\odot$, obtaining $\log(F_{\mathrm{H}\alpha}/f_{\lambda,\mathrm{FUV}}) \approx 1.14$ (in units of $\log$(\AA)) for a Salpeter IMF. 
They further demonstrated that this ratio decreases to 0.43 for a steeper IMF slope of $\alpha = 3.3$, and to 0.06 when the upper mass limit is reduced to $M_{\mathrm{u}} = 20$\, M$_\odot$, underscoring its sensitivity to the most massive stars. The measured values are broadly consistent with expectations for a standard Salpeter IMF.


In \cite{Amrutha_2025}, the IMF index $\alpha$ was estimated using FUV, NUV, and H$\alpha$ emissions. By using these emissions, the approximate number of O, massive B stars (B0; $21 M_{\odot} \geq M_{*} \geq 10 M_{\odot}$) and low mass B stars (B1; $10 M_{\odot} \geq M_{*} \geq 3 M_{\odot}$)  were estimated for each SFC. Taking the ratio of the number of O to B0 stars and of B0 to B1 stars yielded two IMF indices. However, $\alpha$ obtained using B stars was considered reliable in the study. The equation used in \cite{Amrutha_2025} to estimate IMF index $\alpha$ is,

\begin{equation} \label{eq8}
\frac{N(B0)}{N(B1)} =\frac{m_{u}(B0)^{1-\alpha}-m_{l}(B0)^{1-\alpha}}{m_{u}(B1)^{1-\alpha}-m_{l}(B1)^{1-\alpha}}
\end{equation}

\noindent
Where N(B0) and N(B1) are the estimated numbers of B0 and B1 stars. $m_{u}$ and  $m_{l}$ are the upper and lower mass limits, respectively.

We used H$\alpha$ and GALEX FUV and NUV data to estimate $\alpha$ as UVIT has only FUV observations. We convolved GALEX FUV and H$\alpha$ data to GALEX NUV resolution (5.3$^{\prime\prime}$), the poorest resolution of all other observations. We extracted SFCs from GALEX FUV and got the flux for each SFC. We used the parameters of these SFCs and overplotted them on other bands to get the flux from them. We corrected luminosities of SFCs for internal extinction using the extinction maps as described in section \ref{sec:sfr_estimation}. We further used the method outlined in \cite{Amrutha_2025} to get the $\alpha$ estimate from the B stars. For this estimation, we assumed the H$\alpha$ emission comes from O8V stars, as it was shown that the $\alpha$ value estimated from B star ratios does not depend much on the type of O stars considered.

The median IMF slope is $\alpha$ = 1.74 $\pm$ 0.08 in the outer disk and $\alpha$ = 2.45 $\pm$ 0.06 in the inner disk (Figure~\ref{fig:IMF}, left panel). The shallower slope in the outer disk implies a higher proportion of massive stars. While over 50\% of the SFCs across the disk have $\alpha$ values between 2 and 3, several outer-disk SFCs display a top-heavy IMF, indicating an enhanced population of massive stars and suggesting that star formation in these regions is relatively recent.

The elevated $\log(F_{\mathrm{H}\alpha}/f_{\nu,\mathrm{FUV}})$ in the outer disk, together with the comparatively shallower IMF slope ($\alpha$) relative to the inner disk, indicates that massive-star formation remains efficient even in these low-density regions. This suggests that the outer disk of extended UV galaxies can form a relatively larger fraction of high-mass stars, implying that the IMF there is not truncated and may be slightly top-heavy. Outer disks are also metal-poor compared to inner regions, and metallicity plays a critical role in the thermal evolution of molecular clouds: higher metal abundances enhance cooling leading to lower gas temperatures, and reduce the characteristic stellar mass. Consequently, metallicity is considered a key factor influencing IMF variations \citep{Omukai2000ApJ...534..809O, Omukai2005ApJ...626..627O, Schneider2006MNRAS.369.1437S, Schneider2012MNRAS.419.1566S, Chiaki2014MNRAS.439.3121C}. Recent three-dimensional simulations further demonstrate that cloud fragmentation occurs once a finite metallicity is reached \citep{Clark2008ApJ...672..757C, Dopcke2011ApJ...729L...3D, Dopcke2013ApJ...766..103D, Chiaki2016MNRAS.463.2781C, Safranek2016MNRAS.455.3288S}, leading to the formation of low-mass stars due to dust-induced cooling. In the metal-poor outer disk of NGC\,2090, inefficient cooling may suppress fragmentation, favoring the formation of massive stars and naturally producing a top-heavy IMF. These findings underscore how environmental conditions, particularly metallicity and dust content, can modulate the stellar mass distribution in extended disks and influence galaxy-wide star formation processes.

\section{Summary}
The multiwavelength analysis demonstrates that NGC\,2090 is undergoing significant star formation in its extended outer disk. The detection of both FUV and H$\alpha$ emission at large galactocentric radii confirms the presence of young, massive stars well beyond the central regions, while the inner disk is dominated by redder K-band emission characteristic of an older stellar population. This systematic radial gradient in stellar populations, together with the elevated specific star formation rate in the outskirts, is consistent with inside-out disk growth, in which recent star formation increasingly contributes to stellar mass assembly at larger radii.

The FUV emission extends to $\sim$30 kpc, well beyond the optical disk traced by the g band, placing NGC\,2090 among Type~2 extended ultraviolet (XUV) disk galaxies, which are characterized by blue FUV-NIR colors in optically low surface brightness outer regions \citep{Thilker2007ApJS..173..538T}. Unlike the typically diffuse UV emission seen in many Type~2 XUV disks, NGC\,2090 exhibits structured FUV emission and localized H$\alpha$ regions that follow spiral features in the outer disk. This spatial correspondence suggests that large-scale disk instabilities, such as spiral density waves, can propagate into the low-density outskirts and locally enhance gas densities, thereby triggering star formation even at large galactocentric distances.

In nearby spirals, the stellar disk in addition to gas disk  sets the local instability conditions \citep{Romeo2017MNRAS.469..286R}, with stellar self-gravity triggering perturbations that couple to the gas and promote cloud fragmentation. In NGC 2090, however, the stellar disk is confined to $\sim$5 kpc, while star formation extends to $\sim$30 kpc. This indicates that star formation in the outer disk cannot be driven by stellar-dominated instabilities, but instead likely arises from including local gravitational instabilities in a low-surface-density HI disk, turbulence-driven compression, spiral density waves perturbations, or the maintenance of marginal instability through external gas accretion. \citet{Bush2008ApJ...683L..13B, Bush2010ApJ...713..780B} used hydrodynamical simulations to demonstrate that spiral density waves originating in the inner disk can propagate into the outer gaseous disk, locally enhancing gas densities above the star formation threshold and producing star-forming regions in the extended disk. While their models successfully reproduce Type 1 XUV morphology, they fail to account for the more extended, diffuse star-forming regions seen in Type 2 XUV disks. The latter likely require additional processes such as recent gas accretion, which can replenish the outer disk with low-metallicity gas, creating conditions conducive to in-situ star formation. NGC\,2090, is a type 2 XUV and shows star formation associated with the spiral arm. This suggests that both gas accretion and spiral density wave propagation may be at play: gas accretion could supply the necessary fuel, while spiral density waves could generate localized overdensities that trigger star formation in the extended, low-density outer disk.

The more confined distribution of H$\alpha$ compared to FUV likely reflects the shorter lifetimes of ionizing OB stars, as well as the potential effects of stochastic star formation or reduced ionizing photon escape in low-density environments \citep{Boissier2007ApJS..173..524B, Werk2010AJ....139..279W}. The presence of young stars in the outer disk, where the gas surface density, metallicity, and dust content are typically low, implies that star formation can proceed under suboptimal conditions. Cold gas accretion, either via minor interactions or inflow from the intergalactic medium, may supply fresh gas to the outskirts. As the accreted gas cools and becomes gravitationally unstable, spiral arm perturbations can induce local overdensities that initiate star formation, even in the LSB regions \citep{Keres2005MNRAS.363....2K, Ocvirk2008MNRAS.390.1326O, Brooks2009ApJ...694..396B, Nelson2013MNRAS.429.3353N}. 

The spatial distribution of PAH emission and the 21\,$\mu$m dust emission shows a strong correlation with the FUV emission from SFCs, indicating a close association between dust heating, PAH excitation, and recent star formation activity. Notably, the 3.3\,$\mu$m and 7.7\,$\mu$m PAH features exhibit enhanced emission along the spiral arms, suggesting that PAH carriers are either more abundant or more efficiently excited in these regions. This enhancement likely reflects the influence of density waves and local UV radiation fields that are particularly intense along spiral arms, promoting both star formation and PAH emission.

The observed number and flux ratios of H$\alpha$ to FUV emission in the outer disk of NGC\,2090 indicate that the initial mass function is not truncated in the outer disk. This suggests that the high-mass end of the IMF remains fully populated even in the diffuse, low-density outskirts of the galaxy. The detection of significant H$\alpha$ emission confirms the presence of ionizing O-type stars, which would be absent under a truncated IMF scenario. These results imply that star formation in the extended outer disk proceeds efficiently, sustaining massive-star formation despite the low gas densities and metallicities characteristic of these regions. These observations support the picture of NGC\,2090 undergoing inside-out stellar disk growth. The coexistence of an evolved stellar population in the center and active star formation in the outskirts is consistent with results from resolved stellar population studies of nearby galaxies, which find systematically younger stars at larger galactocentric distances \citep{Williams2009ApJ...695L..15W}. Such extended disk star formation plays a key role in the ongoing mass assembly and morphological evolution of late-type spiral galaxies.

\begin{acknowledgements}
We thank the anonymous referee for their constructive comments, which have significantly improved the clarity and impact of this work. JY acknowledges financial support from the Spanish Ministry of Science, Innovation and Universities (Ministerio de Ciencia, Innovación y Universidades, MICIU) , project PID2022-136598NBC31 (ESTALLIDOS8) by MCIN/AEI/10.13039/501100011033. DR and MD acknowledge the support of "The Royal Society Yusuf Hamied International Exchange Award” funded by The Yusuf and Farida Hamied Foundation, U.K. MD also gratefully acknowledges the support of the Department of Science and Technology (DST) grant DST/WIDUSHI-A/PM/PM/2023/25(G) for this research. This publication uses data from UVIT, which is part of the AstroSat mission of the Indian Space Research Organisation (ISRO) and is archived at the Indian Space Science Data Centre (ISSDC). We gratefully thank all the members of various teams for supporting the project from the early stages of design to launch and observations in orbit. This work is based on observations made with the NASA/ESA/CSA JWST. The data were obtained from the Mikulski Archive for Space Telescopes at the Space Telescope Science Institute, which is operated by the Association of Universities for Research in Astronomy, Inc., under NASA contract NAS 5-03127. The observations are associated with the JWST program 3707. 
    
\end{acknowledgements}
\bibliographystyle{aa} 
\bibliography{references}

\begin{onecolumn}
\begin{appendix}

\section{Supplementary figure}

\begin{figure}[h]
    \centering
    \includegraphics[width=0.99\textwidth]{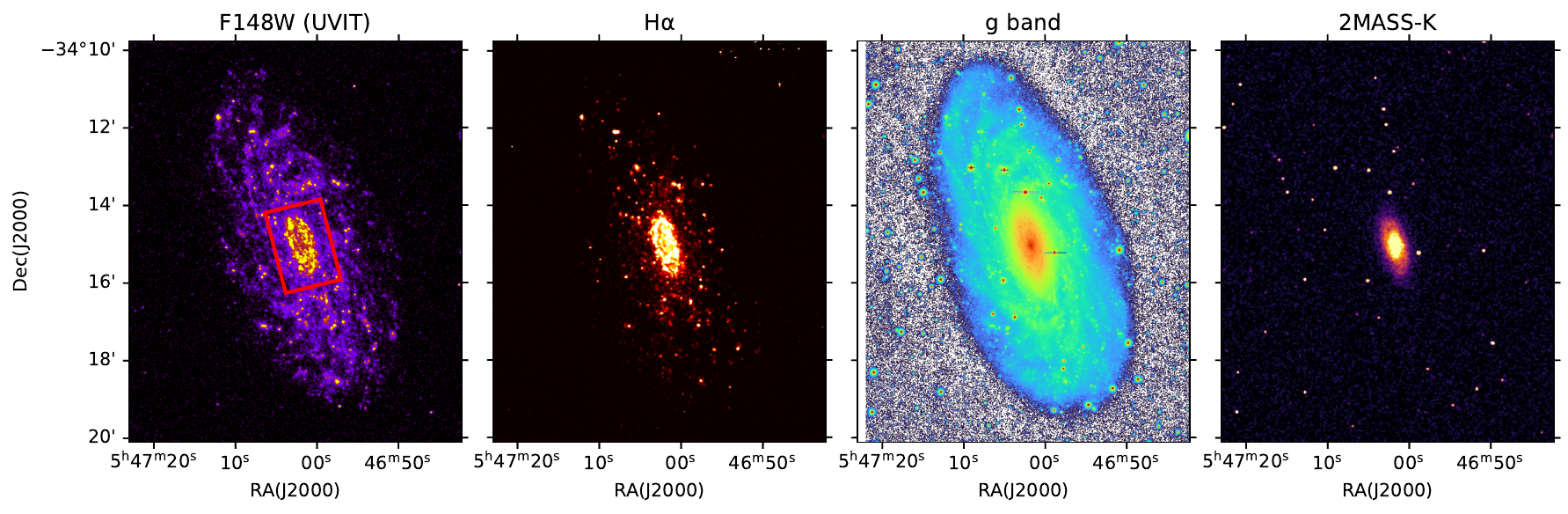}
    \includegraphics[width=0.99\textwidth]{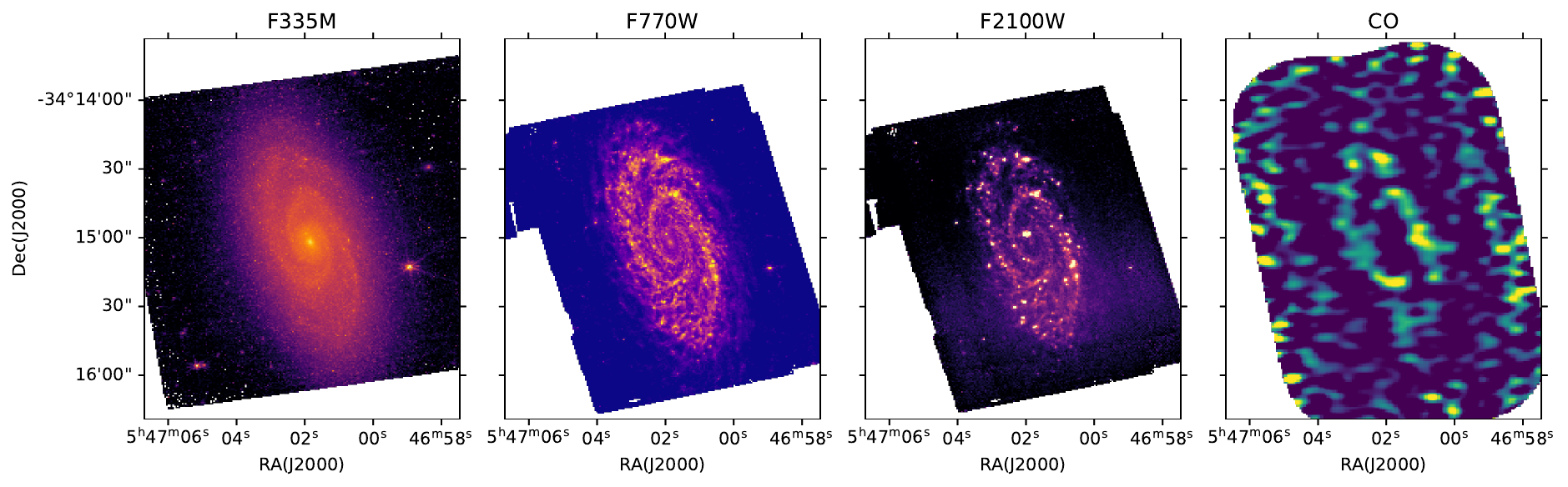}
    \caption{Multiwavelength view of NGC\,2090. Top panel: FUV, H$\alpha$, optical g-band, and near-infrared 2MASS K-band images, tracing young massive stars and the underlying older stellar population. The red rectangle shows the JWST field of view. Bottom panel: JWST images in the F335M, F770W, and F2100W bands and CO map, highlighting PAH emission, warm dust continuum, and molecular gas, respectively.}
    \label{fig:multiwavelength_images}
\end{figure}

\end{appendix}
\end{onecolumn}

\end{document}